# Laser Scanning Microscopy of HTS Films and Devices

A. P. Zhuravel, A.G. Sivakov, O.G. Turutanov and A.N. Omelyanchouk
*B. I. Verkin Institute for Low Temperature Physics & Engineering, National Academy of Sciences of Ukraine, 61103 Kharkov, Ukraine*

Steven M. Anlage
*Center for Superconductivity Research, Department of Physics, University of Maryland, College Park, MD 20742-4111 USA*

A.V. Ustinov
*Physics Institute III, University of Erlangen-Nurnberg, D-91058, Erlangen, Germany*

**Abstract**. The work describes the capabilities of Laser Scanning Microscopy (LSM) as a spatially-resolved method of testing high–$T_c$ materials and devices. The earlier results obtained by the authors are briefly reviewed. Some novel applications of the LSM are illustrated, including imaging the HTS responses in rf mode, probing the superconducting properties of HTS single crystals, development of two-beam laser scanning microscopy. The existence of the phase slip lines mechanism of resistivity in HTS materials is proven by LSM imaging.

## 1. INTRODUCTION

Several review papers dedicated to the applications of the Laser Scanning Microscopy (LSM) in medicine, biology [1,2] and material science [3] are now available. There has been a marked recent upturn in the number of publications on using the LSM as a powerful tool for non-destructive testing of planar microelectronics chips [4-8] containing semiconductor- and normal-metal elements. In our opinion, however, there is a gap to be filled in discussing the exploitation of capabilities of the LSM in experimental physics and technology to study superconductors including high temperature superconductors (HTS). This work presents a review on various applications of low-temperature laser scanning microscopy developed to probe the optical, structural and electronic properties of HTS materials and cryoelectronic HTS devices. We focus on this technique since the LSM capabilities are more than adequate for the applied and research problems in the field. Firstly, the micron-range resolution of the LSM matches the scale of point and extended defects, and HTS material grains. These grains are responsible for the spread of superconducting properties (critical temperature $T_c$, superconducting transition width $\Delta T_c$, critical current density $j_c$) and for the topology of dc and rf electric transport. Secondly, the large field of view provided by the LSM (up to tens of millimeters) is comparable to typical sizes of experimental HTS structures and HTS microchips. Last, the LSM presently seems to be the only technique which gives the option of simultaneous imaging and local manipulation of the superconducting properties of HTSs in a controllable way.

To our knowledge, there are two more low-temperature techniques comparable with the LSM in terms of spatial resolution and imaging contrast. One of them is magnetooptics (MO) which exploits the Faraday effect of rotating the polarization plane of a light beam to explore the variations of magnetic field near a superconductor surface [9]. However, the MO technique is of limited utility for testing superconductors due to its low magnetic sensitivity and low-frequency operation range. This prevents it from imaging the electric properties of HTSs associated with, e.g. the topology of weak rf and dc current flow. Besides, the MO images of spatial magnetic field distributions need an additional computer processing based on complicated mathematical models (such as the inverse Biot-Savart problem) to translate them into transport current maps. The essential limitation of the MO technique as a method of surface analysis is its failure to distinguish between the superconducting properties of different layers in a multilayer HTS structure owing to magnetic screening by the topmost superconducting layer. The LSM is able to illuminate a HTS material through to the optical absorption depth α, which is typically 60 nm to 90 nm for various HTSs for the quantum energy $h\nu_{laser}$ being about 1-2 eV [10].

Another well-developed technique to study spatially the superconducting properties of HTSs is Scanning Electron Microscopy (SEM) [11]. The principal imaging modes used to visualize the 2-D non-uniform distributions of electronic and thermal properties of HTSs are essentially the same for both methods [12]. Also, they have the same spatial resolution governed by the thermal healing length in the HTS object rather than by the sharpness of the probe focusing. Nevertheless, the LSM has some advantages as compared to the SEM, namely:

1) The photons in LSM are electrically neutral. This avoids surface charging effects, which is a major difficulty in the SEM technique.

2) The photon energy in LSM is less than that of the electrons emitted from the SEM gun by at least 3 orders of magnitude. The LSM scanning is therefore more delicate since it does not modify the HTS structure (e.g., does not cause surface oxygen diffusion).

3) Unlike SEM, the LSM studies can be fulfilled in any magnetic environment, from zero field to very strong ones.

4) LSM gives the option of introducing any number of additional laser beams to provide multi-probe measurements.

We have shown earlier that the use of the LSM allows confident search and identification of microscopic defects [13-15], individual grain boundaries [16,17], spatial irregularities of HTS superconducting properties [18,19], and resistive regions of Ohmic dissipation [20]. Additionally, the LSM was applied to analyze the spatial dynamics of resistive properties of HTSs [21] as a function of varying temperature, magnetic field, dc transport current and electromagnetic irradiation. In this review paper, we will illustrate some of the results obtained. Also, we will focus on the newly realized LSM possibilities such as using the rf imaging mode to picture the linear and non-linear HTS responses, visualizing thermoelectric effects, probing the superconducting properties of HTS single crystals, and the development of two-beam microscopy as well.

## 2. Experimental

### 2.1. Principle

In contrast with common light microscopy, the LSM method utilizes point-by-point raster scanning of the surface of a sample-under-test by employing a sharply focused laser beam as the optical probe. The interaction between the light and the superconductor results in local heating of the superconductor by the absorbed laser power, creating a non-equilibrium perturbation in its electronic system, and change in the intensity and polarization of the reflected beam. These changes are the source of the LSM photoresponse PR($x,y$) that is detected as a function of the probe position ($x,y$) within the sample area. The photoresponse is transformed then into the local LSM contrast voltage $\delta V(x,y)$ to put it into computer for building a 2-D image of the optical, electronic, rf and other properties of the superconductor. The laser power is chosen low enough to consider the perturbation small within the framework of the specific task. The probe intensity is modulated in amplitude with the frequency $f_M = \omega_M/2\pi$ and lock-in detected to enhance the signal-to-noise ratio and hence the contrast. The lock-in detection of the ac component $\delta V_{AC} = 0.5 \delta V(x,y,\omega_M) \sin(\omega_M t + k)$ at the frequency $f_M$ is used ($t$ is time and $k$ is phase shift of the excitation wave).

### 2.2. Setup

A simple schematic of one of a series of Low Temperature (LT) LSM arrangements developed by us is shown in Fig. 1. The LT LSM is built up from *(i)* an optomechanical module, *(ii)* a cryostat, and *(iii)* measuring, controlling, and processing electronic units.

The optomechanical module is used to create a nearly diffraction-limited size laser probe, to position it precisely in the area of the sample, and to examine the optical quality of the sample surface using a high-performance (500$^x$) light-microscope operation mode. To form a Gaussian light probe on the sample surface, the laser beam travels from the diode laser (wavelength $\lambda_{opt}$ = 670 nm, maximum

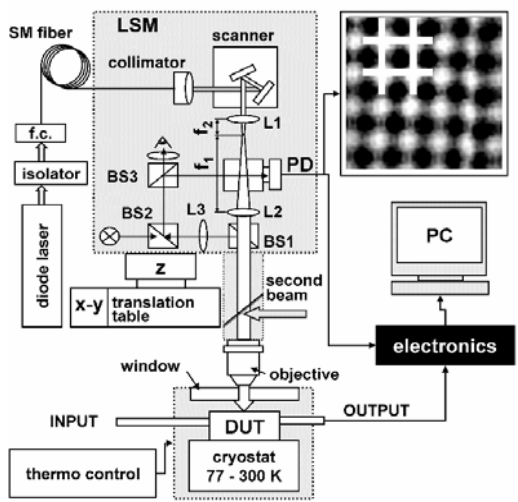

**Fig. 1**. Schematic diagram of the LT LSM optics and cryogenics

**Notations**: L - lens; BS - beam splitter; f - focal length; SM - single mode; f.c. –fiber coupler; PD - photodiode. The inset illustrates the reflective 10x10 μm$^2$ LSM image of the SEM grid with 1x1 μm$^2$ square-shaped through-holes cells separated by 1 μm$^2$ metal strips.

output power $P_{laser}$ = 7 mW), passes through an optical isolator combined from $\lambda/2 + \lambda/4$ wave plates, and then through the fiber coupler to be spatially filtered in the $d_k$ = 4 μm core of a single-mode optical fiber with a numerical aperture (N.A.) of about 0.11. The collimator adapts the laser light from the fiber to the diameter of the scanning module. The transmitted laser radiation then is expanded by a factor M = $5^x = f_2/f_1$ by a pair of the relay lenses ($L_1$ and $L_2$), and passed through the beam splitter ($BS_1$) to fill the entrance pupil of the LSM objective.

In order to scan the probe over the sample in a raster pattern, (*x*) horizontal and (*y*) vertical (orthogonally crossed) scanning mirrors, driven by a pair of galvano-scanners, are assembled in the plane conjugate to the entrance pupil of the objective lens. Typically, a $20^x$, N.A. = 0.42 microscope objective was used to provide fast and accurate laser scanning of samples within a raster area of 250x250 μm$^2$, with a (1/e$^2$) diameter of the beam of 1.2 μm yielding a maximum heating of not more then 1 K while the resulting intensity of the laser probe did not exceed $10^8$ W/m$^2$ on the sample surface. To scan the larger, 1x1 mm$^2$, area, the $5^x$, N.A. = 0.14 objective is also used giving a 5 μm-diameter laser probe. The raster center on the sample can be set by stepper motors in the range as large as 25x25 mm$^2$ with accuracy of 0.5 μm.

The estimation of the probe irradiance and the optical LSM resolution, as well as the calibration of the raster size, was provided in the reflective imaging LSM mode by scanning the SEM grids of different cell size varying from 1x1 μm$^2$ to 150x150 μm$^2$. As an example, a 10x10 μm$^2$ reflective LSM image is shown in the insert to Fig. 1. The image was detected by a photo-diode built in to the optical train of one of the two similar visual channels.

Note that our LSM has several unique features that distinguish it from similar laser scanning techniques described in the literature. This microscope is equipped with additional (not shown) optics to arrange a second scanning laser probe for local irradiation of the sample. For this purpose, a second laser ($\lambda_{opt}$ = 670 nm, $P_{laser}$ = 10 mW) with a structured beam is focused on the sample surface through the same objective lens. Beam-splitting ($BS_1$ - $BS_3$), scanning, as well as relay ($L_2$) optics of the second LSM channel have been created to independently focus and position both laser beams on the sample. The purpose of this independently scanning probe is to manipulate the thermo- and light- sensitive properties of HTS films (such as critical current density, magnetic penetration length and surface resistance) locally. This procedure enables one to artificially model faults, defects and inhomogeneities in HTS structures in order to study their influence on the *local* and *global* characteristics of the sample.

For cooling of the HTS samples into the superconducting state, the LSM was equipped with a home-made, miniature cryostat having a 25 mm diameter glass window to introduce the laser beams and a number of microwave as well as DC electrical feed throughs to connect the sample to measuring electronics. This cryostat stabilized the temperature $T_B$ of the sample in the range from 78 K to $T_C$ with an accuracy of 5 mK by using a resistive heater bifilarly coiled around a copper cold stage. The samples were glued to this stage inside a vacuum cavity of the cryostat.

A number of specific schemes for the LSM electronics for different detection modes have been published elsewhere [13-26] and are not a subject for discussion in the present publication.

### 2.3. Imaging modes

Since the LSM image is formed as a result of an interaction between the laser irradiation and the structure-under-test, any property of the superconductor or the beam which changes during this interaction can serve as the detected photoresponse signal PR(*x,y*). More then ten different modes of detecting the PR(*x,y*) signal are used to create the LSM image contrast while examining various properties of the HTS samples. The exploitation of certain contrast mode (or combination of different ones) is dictated by the problem to be solved. Below one can find a description of the most typical modes, which we have used to obtain the results discussed in this work. The capabilities of the rest, and other promising ones, will be briefly mentioned in the Conclusion.

#### 2.3.1. Optical contrast

The **optical contrast** is intended for visualizing optically resolved irregularities on the HTS structure surface. It appears when registering the power and (or) the depolarization angle of the reflected laser beam during scanning. The optical contrast mode helps in finding surface microdefects, grain boundaries, regions differing by oxygen content in the HTS material and in associating them with the images taken in other contrast modes to establish a correlation between different "visualized" properties. Also, this mode is useful to calibrate the beam intensity, the raster size and to estimate the raster distortion.

*2.3.2. dc voltage contrast*

The **dc voltage contrast** lets one explore the spatial structure of the resistive state of a superconductor. To build the LSM image in this mode, the sample is biased with a dc transport current of value $I_B$ that drives the sample into the resistive (non-superconducting) state under joint action of this current and/or the laser probe. The variation of the voltage generated in the current-biased sample due to its local illumination serves as the response signal to generate LSM contrast.

Phenomenologically, the total change in voltage induced in a small piece of the superconductor can be expressed as

$$dV = \frac{\partial V}{\partial T}dT + \frac{\partial V}{\partial I}dI , \qquad (1)$$

where the first term represents the thermal response mode referred to as the bolometric mode, the second term represents the current response mode referred to as the non-bolometric mode, and $V$ is the detected voltage, $T$ is the temperature, and $I$ is the bias current. To show the dependences of the response voltage on incident power $P$, sample resistance $R$, bias current $I_b$, effective thermal conductance $G(T)$, superconducting gap $\Delta$, and critical current $I_c$ in more explicit form, formula (1) is rewritten as [27]

$$\frac{dV}{dP} = I_b \frac{\partial R}{\partial T}\frac{1}{G(T)} + \frac{\partial V}{\partial I_c}\frac{\partial I_c}{\partial \Delta}\frac{\partial \Delta}{\partial P} \qquad (2)$$

The bolometric part of the response (2) represented by the first term is well known and studied including the case when the incident laser power $P$ is modulated at a certain frequency $f_M$: $P = P_0 \sin 2\pi f_M t$. In most LSM experiments on imaging the temperature-dependent superconducting parameters it plays the main role. However, taking in mind that the bolometric response decays as $1/f_M$, one can vary its contribution in total response by changing the modulation frequency, from an almost purely bolometric mode to the point when the second term starts to prevail. Also, the spatial region where the temperature oscillates due to the modulated laser probe shrinks as the modulation frequency rises [28]. The non-bolometric mode is interesting when studying specific direct effects of the optical irradiation on superconductivity and also to enhance the spatial resolution up to the size of the optical probe. Note that these specific mechanisms of direct change of the superconducting gap $\Delta$ due to incident optical power are described by the third factor $\partial\Delta/\partial P$ in the non-bolometric term in formula (2). The first and the second factors are attributed to the type of resistivity realized in the HTS and can be evaluated from model considerations.

The question of the mechanisms of generation of PR(*x,y*) in HTSs is considered in a number of review papers, e.g. [29-33].

*2.3.3. Thermoelectric imaging mode*

In thermoelectric (both amplitude and phase) imaging (TEI) LSM mode, the HTS sample is not electrically biased. The PR(*x,y*) signals induced by the heating effect of the laser probe generate LSM contrast related to the flow of thermal energy through the HTS film in the normal-conducting state at $T_B>T_c$. [34-36]. In the case of predominantly orthogonal heat diffusion into the substrate, the periodic heating of the film surface by the laser probe produces the temperature difference $\Delta T$ between the film top and the film bottom, exciting a longitudinal thermoelectric voltage because of anisotropy of thermoelectric properties of HTS films. It was shown that the PR(*x,y*) are proportional to the tilt angle α between the crystallographic *c*-axis of the film and the normal to its surface due to the tensor Seebeck effect [37]:

$$\delta V_S(x,y) = \Delta T \Delta S(l_T / d_f)\sin 2\alpha \qquad (3),$$

where $l_T$ is the radius of the thermal spot coinciding with the thermal diffusion length in HTS film; $d_f$ is the film thickness, $\Delta S = S_{ab} - S_c$ is the difference between the value of the thermopower $S_c$ along the crystallographic *c*-axis and the thermopower $S_{ab}$ in the (*a,b*)- plane [37].

The TEI mode of LSM contrast is helpful for the identification of the individual grains of different crystallographic orientation that are nonuniformly distributed in the HTS film area [17].

*2.3.4. High-frequency LSM PR mode*

A number of the LSM imaging modes are used for non-contact investigation of the spatially inhomogeneous HF transport in passive HTS microwave devices. Most implemented methods are ones for probing *(i)* linear HF and *(ii)* nonlinear (intermodulation) current densities, as well as for visualization of *(iii)* the dominant sources of Ohmic dissipation that are nonuniformly generated by HF fields in thin cross-sections of micro-strip HTS devices such as resonators and filters for mobile, cellular and satellite communications.

For bolometric probing of all the thermosensitive HF properties of the HTS films, a resonant mode of the device is excited by external synthesized signal generator operating in the bandwidth of the microwave transmittance. Typically, the transmitted microwave power, $P_{OUT}(f)$, as a function of driving frequency $f$, can be approximately described in the limit of weak coupling by a Lorentzian curve [38]:

$$P_{OUT}(f) = \frac{P_{IN}}{(1-\frac{f_0}{f})^2 + (\frac{1}{2Q})^2} \quad (4)$$

where $P_{IN}$ is the input HF power, $f_0$ is the resonant frequency, and $Q$ is the resonator quality factor.

Both $Q$ and $f_0$ are temperature dependent since they are related to the surface resistance,

$$R_S = \frac{1}{2}\mu_0^2 \omega^2 \lambda_L^3(T) \sigma_1(T,\omega) \quad (5)$$

and the surface reactance of the HTS structure, respectively,

$$X_S(T,\omega) = \mu_0 \omega \lambda_L(T,\omega), \quad (6)$$

where $\omega = 2\pi f$, $f$ is the driving frequency, $\mu_0 = 4\pi 10^{-7}$ [H/m] is the magnetic permeability of free space, and $\sigma_1$ is the real (normal) part of the complex conductivity, $\lambda_L(T) = \lambda_L(T=0)/\sqrt{1-(T/T_c)^m}$ is the temperature dependent (London) magnetic penetration depth, and $m = 2\text{-}4$ is the exponent in its approximate temperature dependence [39,40].

At any fixed HF frequency $\omega$, the optical absorption of laser energy in the oscillating LSM probe periodically heats the HTS film on the thermal wave length scale, $l_T$, for bolometric shifting of HF transmittance $\delta\|S_{21}(f)\|^2$ of the HTS device. This effect causes the thermo-induced modulation of transmitted HF power

$$\delta P_{OUT}(f) \propto \frac{\partial\|S_{12}(f)\|^2}{\partial T}\delta T \quad (7),$$

which is detected by a spectrum analyzer and is used as a signal of the LSM photoresponse, $\delta V_{HF}(x,y)$, creating local contrast of HF LSM images. It is important to keep in mind that the total $\partial\|S_{12}(f)\|^2$ has contributions from the effects of HF resonant frequency tuning $\delta f_0$ and broadening of the spectrum $\Delta f_{3dB}$. Generally, $\delta f_0$ is associated with the change of kinetic inductance $L_{ki} = \frac{\mu_0}{2}\frac{l}{w_S}\lambda_{eff}(T)$ due to the thermal probe, while $\Delta f_{3dB} \sim \Delta(1/Q)$ is directly related to photo-induced modulation of the inverse Q-factor due to an increase in local Ohmic dissipation. In this case, the probe-induced distortion of $P_{OUT}(f)$ of the resonance line shape can be presented as a sum of resistive $\delta V_{HF}^{RES}(x,y)$ and inductive $\delta V_{HF}^{IND}(x,y)$ components of $P_{OUT}(f)$ that can be obtained through partial derivatives of (4) as:

$$\delta V_{HF}^{RES} \propto P_{IN}\frac{\partial\|S_{12}(f)\|^2}{\partial(1/2Q)}\delta(1/2Q) \quad (8)$$

$$\delta V_{HF}^{IND} \propto P_{IN}\frac{\partial\|S_{12}(f)\|^2}{\partial f_0}\delta f_0 \quad (9)$$

At temperatures $T_B$ well below $T_C$, the (HF current driven) resistive component of the LSM PR can be neglected in high-quality HTS films. Then, only the reactive term $\delta V_{HF}^{IND}(x,y)$ is present to produce LSM contrast [14, 41-43]

$$\delta V_{RF}^{IND}(x,y) \propto -\frac{\partial P_{HF}}{\partial f}f\frac{\mu_0}{2}\frac{\left[\lambda_{eff}(x,y)J_{RF}(x,y)\right]^2}{W}$$
$$Ad_f\left(\frac{\delta\lambda_{eff}(x,y)}{\lambda_{eff}(x,y)}\right)$$
$$(10)$$

due to modulation of kinetic inductance of the device by the thermal probe. This allows one to measure a quantity proportional to $J_{RF}^2(x,y)$ [41]. Here, $\delta\lambda_{eff}$ is the photoinduced change in effective penetration depth for $d \ll \lambda_L$, $\lambda_{eff} = 2\lambda_L^2/d_f$, and $A$ is the area of the thermal spot (related to $l_T$).

In the nonlinear (intermodulation) imaging LSM mode, two fixed frequency signals ($f_1$ and $f_2$) are applied to the HTS resonator. The signals at frequencies $f_1$ and $f_2$ are centered on the $|S_{12}(f)|$ curve with a close spacing $\Delta f$ and have the same amplitude. Changes in $P_{f1}$ or $P_{2f1-f2}$ as a function of position $(x,y)$ of the laser beam perturbation on the sample are imaged. A spectrum analyzer is used to measure the power in the tones ($P_f$) at these intermodulation (IMD) frequencies to see exactly where in the device these tones are generated. The change in IMD transmitted power $P_{2f1-f2}$ is given by:

$$\frac{\delta P_{2f_1-f_2}}{P_{2f_1-f_2}} \sim \begin{cases} \dfrac{\delta\lambda}{\lambda} - \dfrac{\delta J_{IMD}}{J_{IMD}} \\ -\dfrac{\int \delta\lambda\, R_s\, J_{RF}^2\, dS + \int \delta R_s\, \lambda\, J_{RF}^2\, dS}{\int \lambda\, R_s\, J_{RF}^2\, dS} \end{cases} \quad (11)$$

where $R_s$ is the surface resistance, $J_{IMD}$ is the nonlinearity current scale [46], and $\delta J_{IMD}$ is the change in nonlinearity current scale caused by the laser heating. Hence the LSM IMD photoresponse is related to changes in the local nonlinearity current scale as well as changes in penetration depth and surface resistance at the site of the perturbation.

We measure the change in IMD produced by heating a small area of the sample. Presumably, the more nonlinear parts of the material will contribute a bigger change to the IMD power when they are heated. Based on numerical simulations with the two-fluid model, we can assume to first approximation that the contrast seen by the LSM tuned to the intermodulation frequency is proportional to the local change in intermodulation current density scale, $J_{IMD}$.

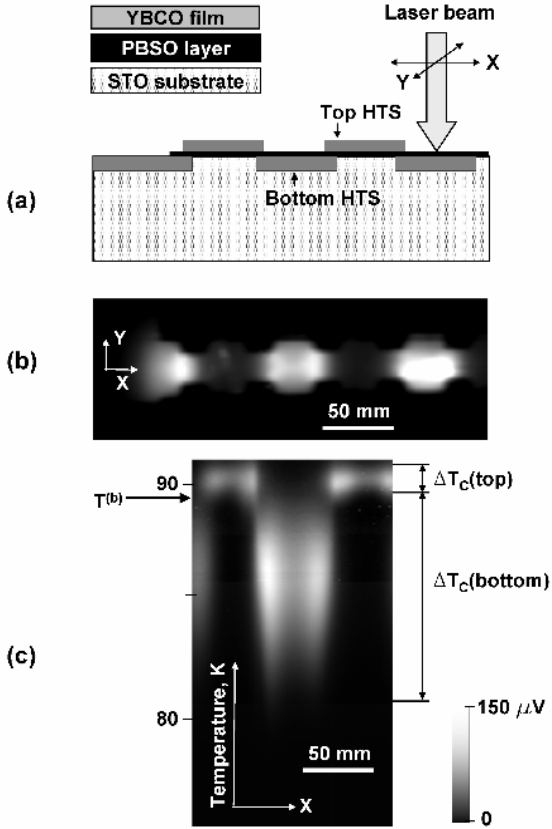

**Fig.2**. Josephson junction array: (a) sample cross section and scanning schematic; (b) half-tone response map in dc voltage mode at $T$=89.5 K; (c) superconducting transition of an array section imaged as temperature series (vertical) of a single scan line (X in (b)).

## 3. EXAMPLES OF USING LSM IN DIFFERENT MODES

Now we shall illustrate the exploitation of the LSM in some of its many possible operating modes.

### 3.1. Using dc voltage and other dc contrast imaging modes

Initially, after the discovery of high-$T_c$ superconductivity the HTS thin films were plagued by low critical parameters. The LSM was used in the dc voltage contrast mode to measure local values of critical temperature $T_c$, transition width $\Delta T_c$, and critical current densities $j_c$ of the films fabricated by various techniques [26,47]. It was shown that the low $T_c$, $j_c$ and broad transitions were caused by local film inhomogeneities rather than the properties of the HTS material itself. As evidence, regions with comparatively high $T_c$'s and small $\Delta T_c$'s were observed in those films [18].

As the HTS fabrication technology progressed, the manufacturing of complicated, multi-element and multi-layer thin film HTS devices began. Therefore, a demand arose for a non-destructive technique for testing the superconducting characteristics of the individual elements of these structures.

Thus, a series Josephson junction array was tested by the LSM [19]. The array cell consisted of a bottom $YBa_2Cu_3O_{7-x}$ (YBCO) thin film electrode deposited onto strontium titanate (STO) substrate, normal $PrBa_2Cu_3O_x$ (PBCO) interlayer and topmost YBCO electrode (Fig. 2,a). The bottom and top electrodes overlapped creating S-N-S Josephson junctions.

The 2-D LSM voltage response map taken at a certain temperature within the superconducting transition of the whole structure (Fig. 2,b) shows that the responses of the top and bottom electrodes are different, indicating a difference in their critical parameters. To determine the critical temperatures and the transition widths of the individual elements, it was sufficient to scan the array along the single mid-lateral line (X in Fig. 2,b) while decreasing the temperature (the response did not change in the transverse direction).

The response $\delta V(X,T)$ as a function of the co-ordinate $X$ and the temperature $T$ in the form of half-tone map (Fig. 2,c) illustrates the transition of an array fragment into the superconducting state. (The brighter regions denote the higher response.) Following the slope of the local transition curve $R(T)$, the response of each element rises as the temperature falls, peaked at a certain temperature and then vanishes to zero. The transition widths of the bottom electrodes are obviously larger than those of the top ones. This simple test can be successfully applied in other cases of quasi-1-D systems where the role of the temperature would be played by another parameter (magnetic field, beam modulation frequency, etc.)

Similarly, a multi-turn thin-film antenna for a HTS SQUID was tested by the LSM in reference [48]. The device included the planar coil, the dielectric layer, the lead-out strip lying over the windings, and a protective STO layer. The critical parameters of the individual components of this multi-layer structure were determined separately. The quality of the *bottom* YBCO layer which had the lower $T_c$ was shown to be the property limiting the critical current rather than the overlap of the thin film strips. With the LSM technique we avoided the difficulties caused by the surface charging effect, which was an issue when using the low temperature electron scanning microscopy to study a similar structure [49].

While it looks reasonable to make a set of electric terminals at specified points of a static structure to make measurements of its specified parts, this would fundamentally disturb the process in spatially-dynamic systems. With the LSM, the propagation of the normal thermal domain (hot spot) in a YBCO thin film was demonstrated in reference [50] (Fig. 3). The edges of the hot spots are always at the temperature $T=T_c$ and therefore give the highest response. Such a voltage-controlled domain can be the basis for a «self-scanning» photodetector or bolometer.

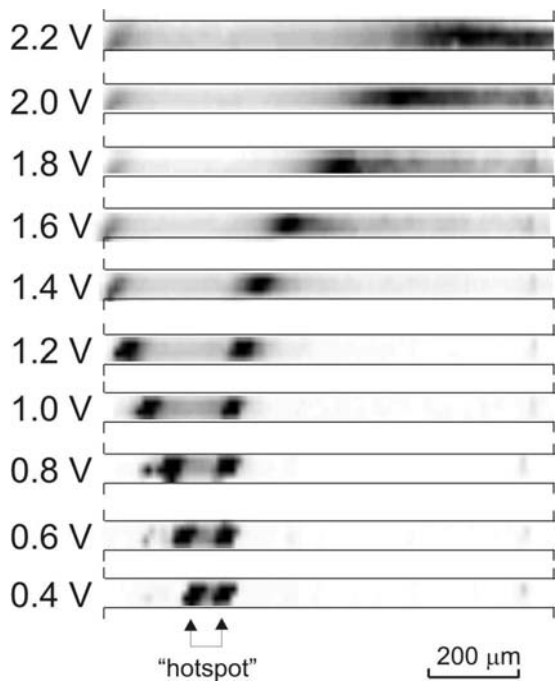

**Fig.3.** Propagation of normal thermal domain (hot spot) in voltage-biased YBCO/STO film strip. The voltage bias value (indicated in figure) determines the hot spot size. Film thickness $d=100$ nm, critical temperature $T_c=85.2$ K, critical current density $j_c(77\text{ K})=2.5\cdot10^5$ A/cm$^2$, temperature $T=76.7$ K, beam modulation frequency $f_M=81.78$ kHz.

One of the global issues researchers face when studying HTS thin films is making a correlation between the superconducting properties (say, critical current density) of these films, and the originating microscopic mechanisms. The modern critical current models of HTS thin films are based on the following facts: (*i*) even epitaxial thin films have micron-size polycrystalline structure; (*ii*) by the results of magneto-optical studies, there is a certain spread in critical parameters among individual grains; (*iii*) the experiments on bi-crystals show that due to crystallographic misorientatoin between the adjacent grains, Josephson weak links are formed at the grain interface, the critical current of which being a function of the misorientation angle, and Josephson vortices flow along the interfaces; (*iv*) the current flow in this irregular structure is of a percolative nature.

Such a model of the polycrystalline film that assumes an exponential dependence of the critical current on grain boundary angle is developed in reference [51]. The 2-D picture representing the rates of magnetic flux flow along the grain boundaries is calculated in [51] by computer simulation. The LSM is able to image this process directly. The LSM dc voltage response map taken experimentally on a polycrystalline thin YBCO film (Fig. 4) strongly resembles the simulated picture (Fig. 6b from [51]). If the transport current slightly exceeds the critical current providing linear vortex flow, then the same map, according to [11], represents also the distribution of the critical current density in the film.

The exploitation of the different modes of LSM contrast in a single experiment allows the researcher to attribute spatial details of resistive

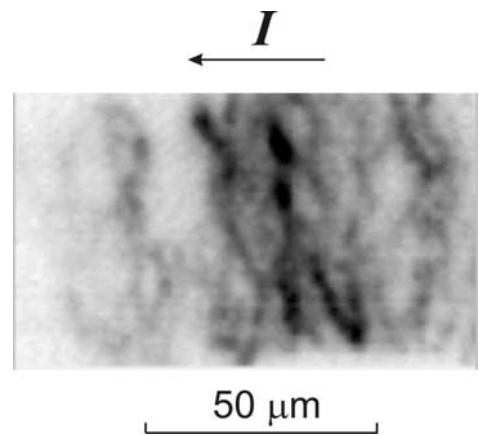

**Fig. 4.** Half-tone dc voltage response map of polycrystalline YBCO film with percolative current flow through a maze of weak links between grains at $T<T_c$. The arrow denotes current flow direction.

processes in the film to its structural features.

Fig. 5,a shows an optical LSM image of a YBCO sample shaped by laser cutting from a 1 μm-thick film deposited on a Ni-W substrate. As seen from this picture, little can be said about the microstructure of the HTS film. Fig. 5,b presents the same part of the sample taken now in thermoelectric LSM contrast at room temperature [17]. This LSM imaging mode based on the Seebeck effect clearly distinguishes between individual grains of the HTS structure, which look like more or less dark spots with typical sizes of about 40 μm. The structure of epitaxially grown YBCO blocks is obviously dictated by the granulated structure of the Ni-W substrate, which shows scattered orientation of the *c* axis. Both theoretical analysis [52] and experimental LSM observations [17] directly indicate that the weakest regions where the superconductivity starts to be destroyed by current are located at the large-angle grain boundaries. This is confirmed by comparison of the LSM images obtained in thermoelectric contrast mode (Fig. 5,b) and in dc voltage contrast (Fig. 5,c). The latter mode visualizes the spatially-nonuniform resistive state. The regions with reduced critical current look like bright lines in Fig. 5,c. Their positions correlate with the positions of weak links formed at the most misoriented grain boundaries in this HTS thin film.

The nucleation of the resistive state of superconductors into individual resistive domains, i.e. phase slip lines (PSLs), and their unusual spatial dynamics as superconductivity is destroyed by current near $T_c$, is a well-established experimental fact by now. It was proven in our previous works on imaging the resistive state of low-$T_c$ superconductors by the LSM [24,53-55]. However, due to the small coherence length, the PSL-based state is difficult to identify in HTS samples whose superconducting properties are extremely sensitive to crystal structure imperfections, defects, and geometric irregularities.

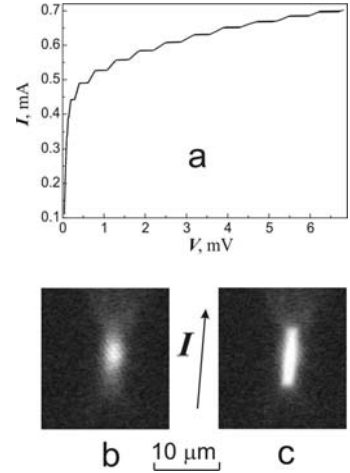

**Fig. 6.** Phase slip lines in a BSCCO single crystal microstrip. (a) I-V curve of 2 μm-wide microstrip, (b) dc voltage response map of a single PSL, (c) the same map taken at a higher current with multiple PSLs filling the whole strip.

Nevertheless, we observed I-V curves with excess current and voltage steps with equidistant differential resistances, that is characteristic of PSLs, in clean defect-free $(Bi,Pb)_2Sr_2Ca_2Cu_3O_x$ (BSCCO) single crystal strips (Fig. 6,a). The strips with dimensions of 10 μm × (1-2) μm × 0.3 μm and edge roughness of 0.01 μm were obtained from the single crystal by ion etching. The PSL size is estimated by the differential resistance of the voltage steps to be 0.2 μm, which is beyond the spatial resolution of the LSM. However, the LSM voltage map obtained at the current value corresponding to appearance of the first voltage jump exhibits a localized formation in the middle of the sample (Fig. 6,b). As the current increases so does the number of PSLs, which then progressively fill the whole sample, while the differential resistance becomes equal to the normal state value.

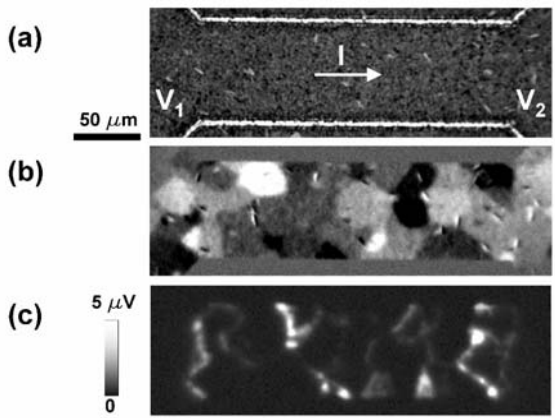

**Fig. 5.** 100x250 μm LSM images showing (a) optical image of HTS sample, (b) thermoelectric map of its granulated structure and (c) resistive response at the grain boundaries.

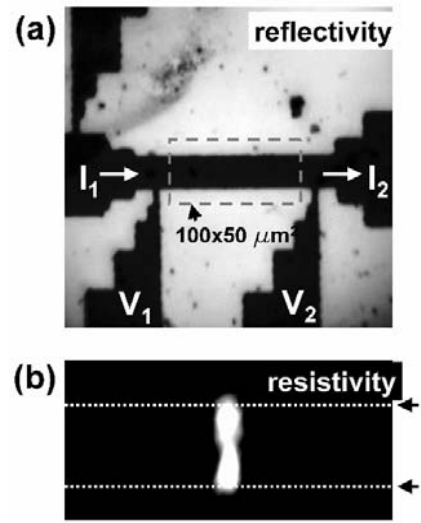

**Fig. 7**. Phase slip lines in YBCO thin film strip. (a) 200x200 μm optical image of the thin film sample, (b) single PSL in the strip center. The YBCO strip is roughly outlined by white dashed line.

After this, the non-zero response becomes uniform over the sample length (Fig. 6,c), in complete agreement with the PSL concept, and excludes the alternative explanations in terms of a thermal domain. In the case of the thermal domain, when the sample resistance is equal to the normal state value, the overheated (above $T_c$) region should occupy the entire sample, and the LSM response would vanish to zero.

A similar picture was observed on "ideal" YBCO thin film samples with $T_c$ = 92.5 K and $\Delta T_c$ = 0.2 K (Fig. 7).

Fig. 7,a displays an optical image of the central part of this HTS structure. It contains a YBCO microbridge 25 μm-wide, 120 μm-long biased by dc current using the terminals $I_1$ and $I_2$ along the direction denoted by arrows. The potential leads $V_1$ and $V_2$ were used to register the response voltage $\delta V(x,y)$ while raster scanning the LSM beam over the area outlined by dark dashed line.

Fig. 7,b illustrates the resistive state of the HTS microbridge biased by the current at which the first voltage jump appears in the sample IV curve. This jump implies the creation of a single PSL whose position is clearly seen on the 2-D voltage response map.

These experiments prove that the phase slip mechanism of resistivity discovered in classic BCS low-$T_c$ superconductors is also valid for the current-induced destruction of superconductivity in HTS materials.

It was experimentally established that in the case of HTSs the LSM can be applied not only to thin films but also to comparatively bulk (several tens of microns in thickness) objects, particularly, to single crystals. Apparently, this is owing to the important role surface pinning of vortices plays in the critical current mechanisms in HTSs. In this case a minor surface perturbation by the laser probe results in a reasonable change in the pinning potential and, hence, leads to fairly detectable alteration of the resistive state of the entire sample.

It is a well-known fact that, as a rule, the almost regular system of defects, the twin boundaries, appears in YBCO single crystals during their growth. The orientation of *a* and *b* axes in the *a-b* plane differs by 90 degrees for the grains separated by a twin boundary. Such a boundary is of atomic scale and can be a pure Josephson weak link since the superconducting order parameter is known to be reduced at the twin boundary. The transport properties of these single crystals are highly anisotropic and depend on the mutual orientation between the current and the twin boundaries.

In addition to the anisotropy, the resistive measurements in single crystals are also complicated due to Joule heating from current feeding terminals. This is because the large critical current densities and large (as compared to a thin film) thickness of the crystal produce high absolute values of the transport current used in the resistive experiments.

Bearing in mind all these considerations, samples having a special geometry were formed in a YBCO single crystal by laser cutting [56] to meet the requirements of the resistive and the LSM studies (Fig. 8). In each strip sample, the twin boundaries had a specified orientation compared to the transport current driven along the sample. Each sample had wide banks to improve heat sinking and to increase the crystal-to-terminal contact area. The narrow potential leads provided for standard four-probe measurements of resistance.

In the case when the twin boundaries are parallel to the current flow direction, they effectively pin the Abrikosov vortices. The critical current is high, and the single crystal behaves like a common classic superconductor demonstrating the Meissner effect and pushing the supercurrent out to the edges (Fig. 9). (The darker areas in Fig. 9 correspond to LSM higher response. It was shown in [11] that the higher thermoinduced response implies higher supercurrent density when measuring the response in the transverse direction.)

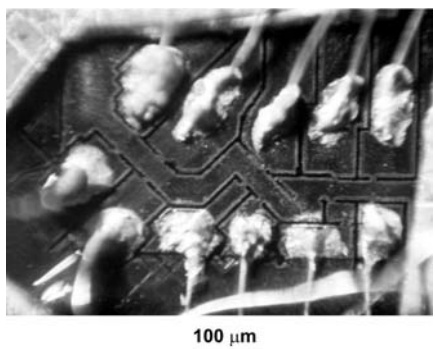

**Fig. 8**. Microphoto of YBCO single crystal samples prepared by laser cutting and oriented in preset crystallographic directions. The electrical contacts to pads are made by conducting paste.

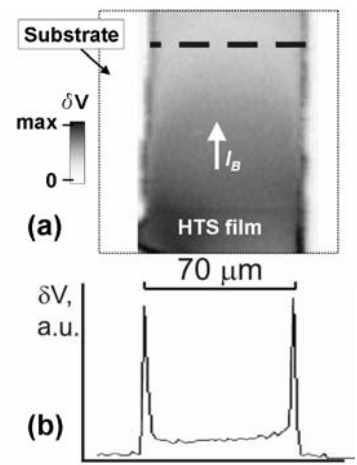

**Fig.9**. Meissner effect in YBCO single crystal containing twin boundaries parallel to current flow direction. (a) Half-tone voltage response map at a current slightly below $I_c$ before scanning, (b) plot of response in single scan line denoted by dashed line in (a).

The current distribution in the single crystal changes dramatically if one drives the current across the twin boundaries (Fig. 10). Unlike the previous case, the supercurrent density peaks in the middle of the sample as the temperature decreases. Earlier, the effect of increasing the density of supercurrent in the middle of a thin film strip was also observed in classic superconductors (tin films) [57]. In that case it was explained by annihilation of self-current vortex-antivortex pairs in the center according to the Aslamazov-Lempitsky theory [58]. A similar effect, according to theory [59], may be evidence for a superlattice of flat Josephson weak links.

The voltage response map obtained on the same sample but with higher spatial resolution (Fig. 11) argues for the latter statement. The appearance of resistivity in localized sites means that some twin boundaries (dark regions) having the lowest critical currents are visualized (within the spatial resolution of the LSM). This image correlates with the picture seen in polarized reflected light, thus confirming that it is the twins that are responsible for the observed resistive structure.

The existence of natural weak links in YBCO single crystals is an alternative option to the creation of artificial Josephson junctions, whose fabrication presently is extremely problematic because of the small coherence length in HTS materials.

Another technically important bulk HTS object that was tested with the LSM was a segment of HTS power cable. The most vital question for the cables is the enhancement of their current-carrying ability and the search for the reasons liming their critical current. To make the LSM measurements, fragments of Ag-coated filaments were extracted from different sections of a BSCCO/Ag tape. The silver was eliminated by laser cutting [56], and then samples having thickness of 10-15 μm and length of up to 2 mm were cut out from the filaments [60].

Fig. 12 demonstrates the LSM voltage response maps of the fragments of superconducting filaments extracted from different parts of the tape cross section. Taking the integral over the distribution of the response voltage in every cross section, one can reduce the 2-D response map to a 1-D problem and determine the critical currents and the critical temperatures for each cross section. Such a procedure for retrieving the spatial distribution of the critical current $I_c(x)$ and the critical temperature $T_c(x)$ along the superconducting filament is described in reference [61].

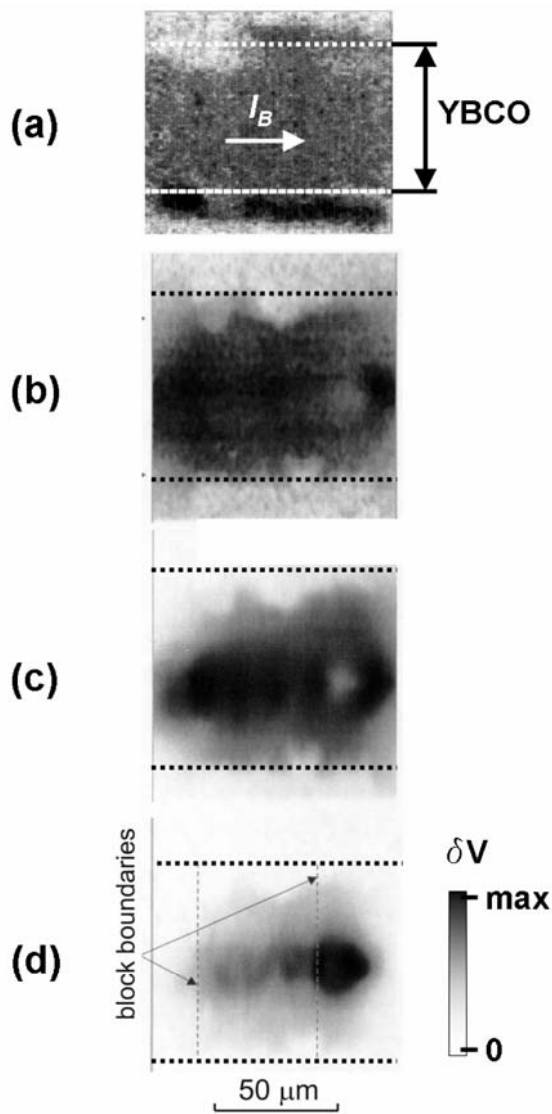

**Fig. 10**. Current flow in the same YBCO single crystal as in previous figure but with twin boundaries across the current direction. Temperature: (a) 91.6 K, (b) 90.6 K, (c) 90.63 K, (d) 90.34 K. Current $I$=10 mA, beam modulation frequency $f_M$=6.79 kHz. Darker areas represent higher response.

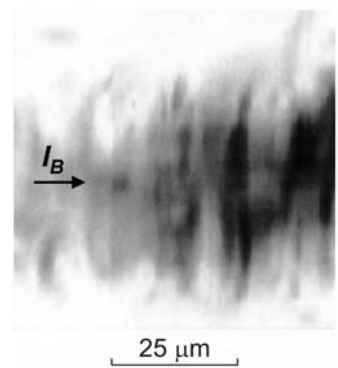

**Fig. 11**. Imaging of twin boundaries (TB) in YBCO single crystal. Current flows across the TBs. Darker areas correspond to higher response voltage. Temperature $T$=90.03 K, current $I$=28 mA, beam modulation frequency $f_M$=43.9 kHz.

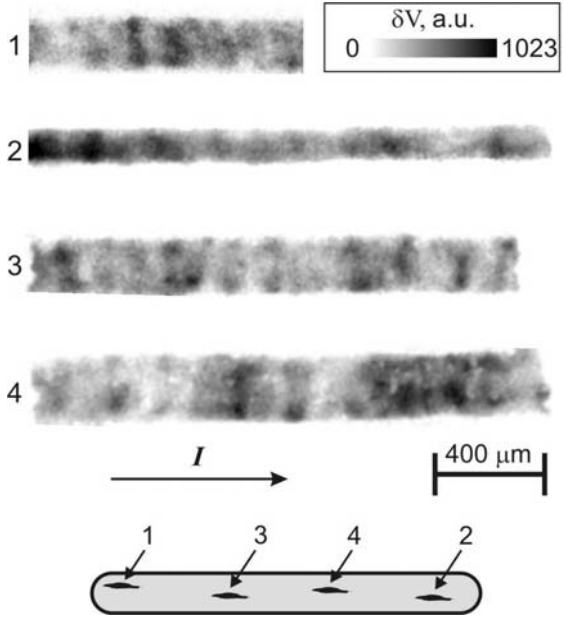

Fig.12. LSM dc voltage response maps of fragments of superconducting filaments extracted from different sites of BSCCO/Ag tape cross section (denoted by numbers in schematic of the tape cross section).

Owing to the LSM, it was found that the current-carrying ability of the filaments from different parts of the tape cross section differ, while the critical current of each filament varies along its length quasi-periodically, reflecting some special features of the fabrication procedure. The LSM studies of the filaments in the non-bolometric mode at high beam modulation frequency has given grounds to consider the appearance of Josephson weak links at the grain boundaries as one of the mechanisms limiting the cable critical current [55].

### 3.2. Examples of HF imaging contrast

In order to illustrate the advantages of the HF (radiofrequency, RF) imaging modes, we have tested the LSM on a specially selected HTS

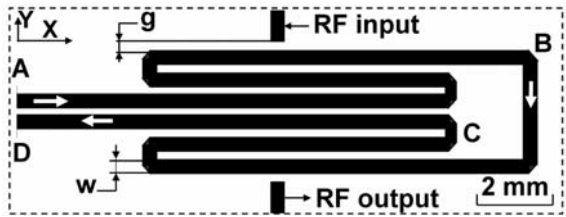

Fig.13. A schematic sketch of the 850 MHz resonator structure used in the measurements. White arrows show the longitudinal path from A to D used in Fig. 16 for plotting the profiles of the standing wave patterns.

resonator containing structural defects and geometrical irregularities. Fig. 13 shows the design of the resonator. The 3 μm-thick superconducting $YBa_2Cu_3O_{7-\delta}$ (YBCO) films were grown on a 0.5 mm-thick $LaAlO_3$ (LAO) substrate. The film exhibits a resistive transition between 89.9 K and 92.2 K as established from transmittance measurements at 1 GHz. The bottom surface of the substrate was glued to a microwave package that serves as a ground plane. Keeping the device structure to a minimum size, the resonator was patterned by ion-milling lithography of the top-surface YBCO film using the meandering strip design. It was rated for a fundamental ($\lambda/2$) resonance frequency, $f_0$, of about 850 MHz with a loaded $Q_L \sim 5670$ at $T = 78$ K. The width of the superconducting strip along the meander line was $w = 250$ μm. The resonator was coupled through two capacitive gaps ($g = 200$ μm) separating the input/output rf electrodes from the rf circuit delivering/measuring power $P_{IN}/P_{OUT}$ in the range from –50 dBm to +10 dBm. All the illustrated LSM experiments were performed at $T_B = 79.5$ K.

Two microwave signals at frequencies $f_1$ and $f_2$ were centered with 0.2 MHz spacing around the resonant frequency of the device, $f_0 = 849.7$ MHz, and injected into the resonator. The nonlinear mixing in the device generates third-order IMD signals at $2f_1 - f_2$ and $2f_2 - f_1$ frequencies. Fig. 14 displays the *global* output spectrum of the resonator. The LSM PR amplitude is measured at either a main tone frequency ($f_1$ or $f_2$) or at an IMD frequency ($2f_1 - f_2$ or $2f_2 - f_1$) by a spectrum analyzer operating in the single-frequency receiver mode. The LSM PR signal imaged through the primary tones $P_{f1}$ or $P_{f2}$ is called the radio-frequency (RF)

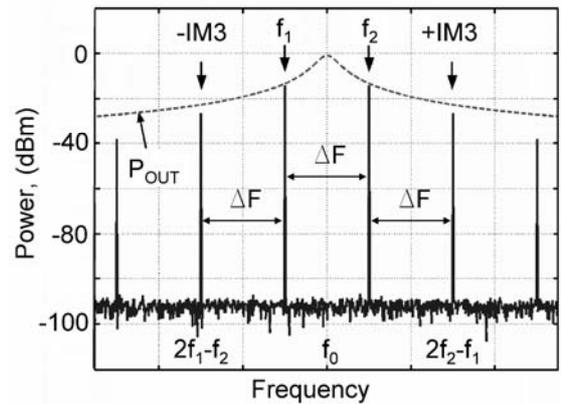

Fig. 14. The resonator (Fig. 14) output spectrum generated as a result of the nonlinear mixing of the two primary tones $P_{f1} = 0$ dBm and $P_{f2} = 0$ dBm at corresponding frequencies $f_1 = 849.6$ MHz and $f_2 = 849.8$ MHz were centered with $\Delta F = 0.2$ MHz spacing around the resonant frequency of the device, $f_0 = 849.7$ MHz. The signals at frequencies $f_1$ and $f_2$ were used for the LSM imaging of both inductive and resistive components of HF PR while the signals at +/- IMD were used for NL sources imaging. The dashed line, $P_{OUT}$, is the linear $|S_{21}(f)|^2$ characteristic measured for calculation of $Q$ and insertion loss.

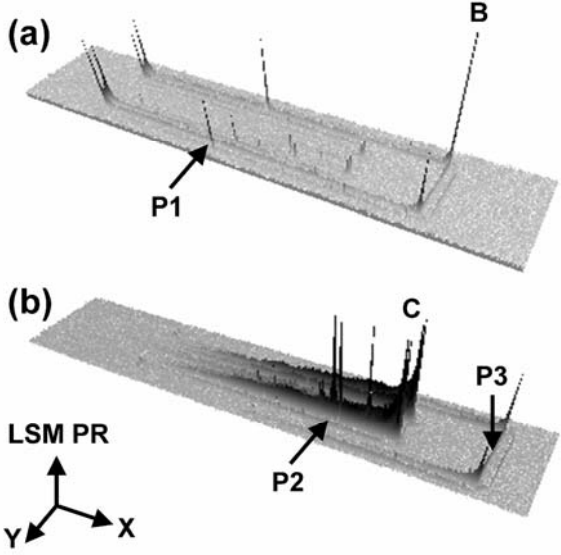

**Fig. 15.** 3D surface of the $\delta V_{RF}^{IND}(x,y)$ distribution generated by (a) the first: 850 MHz and (b) the third: 2.56 GHz harmonic frequency at $P_{IN}$ = 0 dBm. Positions B and C correspond to that shown in Fig.13. Positions P1 and P2 are selected for detailed LSM imaging while P3 is the geometrical center of the YBCO strip-line

PR, while the signal imaged at the intermodulation tones $P_{2f1-f2}$ and $P_{2f2-f1}$ is called the IM PR. The spectral bandwidth of the receiver is about 200 kHz, slightly broader than the modulation frequency $f_M$ = 100 kHz.

Figure 15,a represents a 3-D LSM image showing the spatial variation of RF PR in the (delineated by doted box in Fig. 13) area of the YBCO resonator at its resonance mode. The image was acquired at $P_{IN}$ = 0 dBm and at a frequency $f$ = $f_1$ that is 100 kHz below the fundamental resonance peak ($f_0$ = 849.7 MHz, see Fig. 14). A similar image, probed when the driving signal is at the third-harmonic ($f$ = 2.56 GHz), RF PR image is shown in Fig. 15,b. For clarity, we reconstructed the RF PR(x,y) image into an amplitude profile of $J_{RF}(L)$ along the longitudinal path from A to D (see Fig. 13) constituting the length of the strip, $L_0$ = length of meandering segment D to A. The square root dependence of $J_{RF}(x,y)$ on $\delta V_{RF}^{IND}(x,y)$ (Eq. (10)) was used to calculate $J_{RF}(L)$. To make this profile, data for $J_{RF}(x,y)$ were averaged across the width of the strip conductor for both presentations in Fig. 15. Each plot in Fig. 16 consists of 4000 points separated by 10 µm LSM steps along the L-axis where the ends of the resonator correspond to $L=0$ and $L=L_0$. As evident from the plots, the laterally-averaged HF current density $J(L)$ along the longitudinal direction L of a resonating strip can be represented by a typical standing wave pattern [43-45]

$$J(L) = J_0 * \sin\left(\frac{n\pi L}{L_0}\right), \quad (12)$$

where the peak current density at $L=L_0/2$ is given by [62]

$$J_0 = \frac{1}{wd_f}\sqrt{\frac{r(1-r)8QP_{IN}}{n\pi Z_0}} \quad (13)$$

Here $r$ is voltage insertion loss obtained from the transmission characteristic, $n$ is the index of the harmonic mode, and $Z_0$ is the characteristic impedance of the strip transmission line.

Almost all the $J_{RF}(x,y)$ transverse spatial profiles look like that shown in Fig. 17: a flux-free Meissner state (of middle cross-section P3) having the two peak lines nearly localized at both strip edges [14,41-43]. The exceptions are seen in a few areas marked by the arrows P1, P2, C and B in Figures 13, 15 and 16, where there are a few anomalously high photoresponse values visible due to microscopic defects or superconducting inhomogeneities found in the distributions over the whole resonator topology.

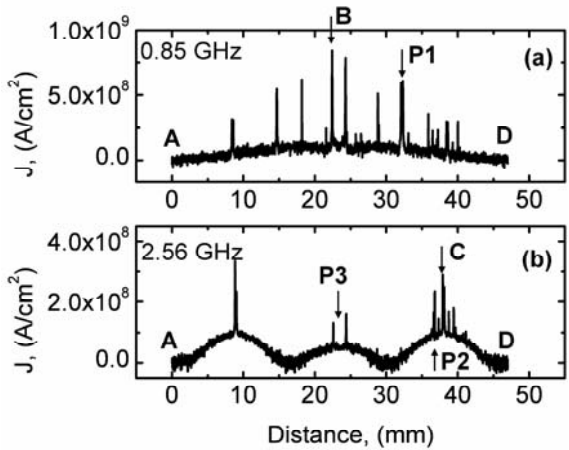

**Fig. 16.** Transverse averaged HF current densities vs. the distance L along the A-D length of the resonator YBCO strip-line showing the standing wave patterns at (a) 850 MHz and (b) 2.56 GHz. The arrows in positions B, C, P1-P3 indicate the particular sections selected for detailed LSM imaging.

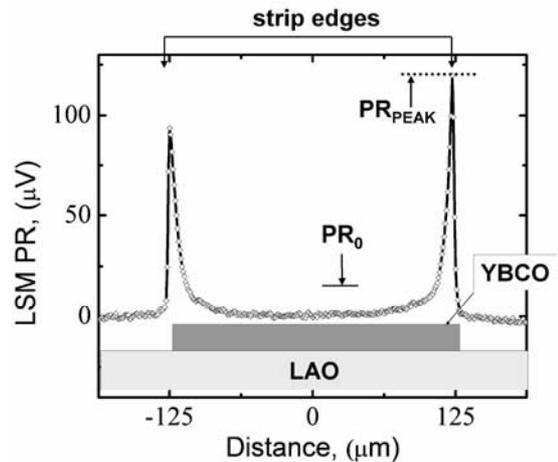

**Fig. 17.** Amplitude profile of the LSM PR distribution in the middle cross-section of the YBCO strip. $PR_0$ denotes the averaged photoresponse while $PR_{PEAK}$ is its maximum amplitude. The section geometry is shown under the profile.

In the middle section P3 (see Figs. 15 and 16), the density of averaged current was about $J(P3) = 6.9 \cdot 10^7$ A/m$^2$ with measured $P_{IN} = 0$ dBm, $r = 8 \cdot 10^{-3}$, $Q = 2800$, and $Z_0 = 47$ Ohm on the third ($n = 3$) harmonic. Analysis of the ratio of the LSM PRs (see profile in Fig. 17) at the edge, PR$_{PEAK}$= 120 to its averaged value, PR$_0$= 8.4, in the section P3 shows that the density of a peak current near the strip edge is about of $J_{PEAK}(P3) = 2.6 \cdot 10^8$ A/m$^2$. This calibration value was used to estimate $J_{RF}(x,y)$ in the other areas of the YBCO structure. For example, the presence of a right angle turn is seen to increase $J_{RF}(x,y)$ by at least one order of magnitude. It is evident from the HF current density spikes in sections B and C corresponding to $J(B) = 1.2 \cdot 10^9$ A/m$^2$ at the first harmonic and to $J(C) = 3.8 \cdot 10^8$ A/m$^2$ at the third one. These anomalies result from the current buildup near the hairpin inner corners that have been theoretically studied by Brandt and Mikitik for dc current density induced by magnetic field in the Meissner-London state [63]. They have shown that the dc current density exhibits a sharp finite peak at the corners; and might enhance the nonlinear Meissner effect. The results of reference [63] can be generalized to the case of $J_{RF}(x,y)$ distribution because of the frequency-independent Meissner screening in superconducting strips.

The power-dependent spatial redistribution of the $J_{RF}(x,y)$ was analyzed in detail in a 50x50 µm$^2$ region positioned not too far from section C. The imaged areas are shown in a reflective 250x250 µm$^2$ LSM image (see Fig. 18,e) by the black dashed boxes centered on two inside corners. The $J_{RF}(x,y)$ undergoes a radical redistribution starting from -12 dBm ($J_{PEAK}(x,y) = 5.4 \cdot 10^8$ A/m$^2$), especially for $P_{IN}$ close to -3 dBm (Fig. 18,a-c). In spite of the fact that the peak value of $J_{RF}(x,y)$ is below the depairing current density (~$3 \cdot 10^{12}$ A/m$^2$ at $T_B = 77$ K in bulk HTS), it can exceed the critical current of "weak" intergrain links in the HTS film.

Fig. 18,c shows the $J_{RF}(x,y)$ distribution when the HF vortices enter YBCO film due to the presence of twin-domain blocks that are formed by the surface topography of the LAO substrate (Fig. 18,e). In contrast, no visible change in the $J_{RF}(x,y)$ map was detected for input powers between -50 dBm and +10 dBm at the left corner. A typical LSM image is presented in Fig. 18,d. The HF vortices here are pinned effectively by the boundaries between the twin-domain blocks that are orthogonally directed to the vortex driving (Lorentz) force.

The presence of such current-blocking obstacles as low-angle GBs (section P1) and microcracks (section P2) gives rise to highly inhomogeneous (often percolative) current distributions in the HTS film [64,65]. From Figs. 15 and 16 it is evident that the LSM PR in those areas is comparable to that produced at the inner corners of the HTS structure. Shown in Fig. 19 are the LSM images for the $J_{RF}(x,y)$ distribution around different kinds of extensive defects crossing the HTS strip.

Fig. 19,a illustrates a non-uniform distribution of $J_{RF}(x,y)$ detected in a 480x320 µm$^2$ area near P1 in the vicinity of a crack in the YBCO film. The crack is formed by an area of sharp misorientation of twin domain blocks in LAO. A

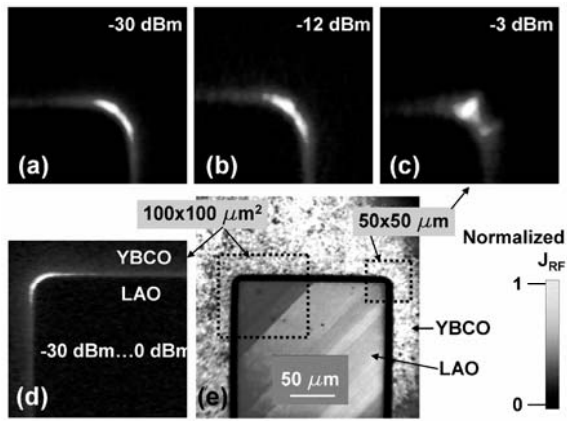

**Fig. 18**. Detailed grey-scale LSM images of the $J_{RF}(x,y)$ spatial variations located near C (Fig. 13) in the area of HTS structure corresponding to (a-c) right-hand side and (d) left-hand inside corners that are crossed by twin-domain blocks of different orientation. Position of LSM scans is shown by dotted boxes in (e) reflective LSM image. Bright (dark) regions correspond to large (small) PR signal. Amplitudes of $J_{RF}(x,y)$ are normalized to get the best contrast.

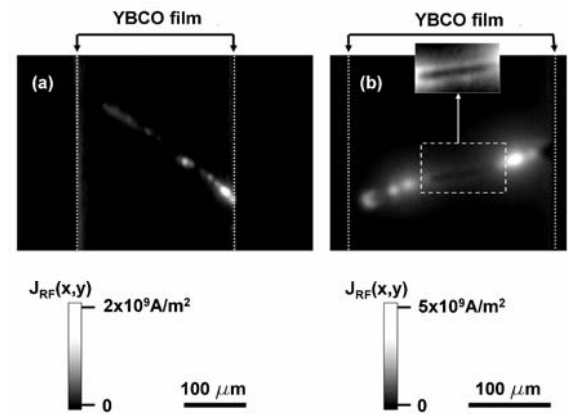

**Fig. 19**. Gray scale representation of the $J_{RF}(x,y)$ distribution in the areas of (a) 480x420 µm$^2$ LSM scan near section P1 and (b) 320x250 µm$^2$ region around the position P2 of Fig. 16. The YBCO strip edges are shown by white dotted lines. The central area of (b) is shown in the inset with enhanced contrast for clarity.

deep penetration of microwave field (that is visible as a spatial modulation of HF current densities) in the HTS strip along the crack may be formed by localized vortices pinned on a twin-domain structure of the YBCO film. It was established that the spatial distribution of LSM PR on this kind of defect is independent of applied HF power in the range from -50 dBm to +10 dBm. This effect does not differ from the effect of the film thinning to increase the effective magnetic penetration depth.

In contrast, the LSM PR undergoes a radical redistribution in region P2 starting from -30 dBm. Fig. 19,b shows LSM in a 300x250 $\mu m^2$ area scan obtained near a linear deep fault there, at the same input power $P_{IN}$ = 0 dBm. The inset indicates the central part of the image plotted with a larger LSM contrast. The dark area between the two bright strips here corresponds to the position of a normal (non superconducting) region of the HTS film resulting from Joule heating within the linear defect. Its behavior as a function of $P_{IN}$ resembles the behavior of a tunnel break-junction structure with a low value of critical current density. At all $P_{IN}$ > -30 dBm, the Cooper pair conductivity is transformed to the conductivity of quasi-particles producing a local overheating effect. This effect gives rise to the resistive component of LSM PR that is visible as a dark area in the contact position. The bright areas in Fig. 19,b determine zones of magnetic penetration on the scale of the normal metal skin depth. The spatial modulation of LSM PR is due to inhomogeneous HF current flow along the contact.

The relation between flux dynamics and pinning processes, both dependent on local electric field, regulates the moment of resistive transition of the HTS film to its overcritical state. In this highly dissipative state, the local RF PR has a more complex nature since it contains an additional resistive component that brings about an error when the LSM PR distribution is converted to $J_{RF}(x,y)$. We eliminate this error in the present work, applying a procedure of processing LSM images for a spatial partition of the HF response correlated with inductive and resistive changes in microwave impedance by measuring the components described by Eqs. (8-9).

Figs. 20, a-c show a spatial modification of resistive HF component of LSM PR$(x,y)$ under the influence of increasing HF power from -12 dBm through -6 dBm to 0 dBm. All the 50x50 $\mu m^2$ images are acquired in region B occupying the inside right-angle turn of the YBCO strip. The cross-section of the patterned micro-strip was an isosceles trapezoid having $60^0$ angles at the bottom of the strip. It is clearly visible as a black YBCO/LAO interface in the reflective LSM image

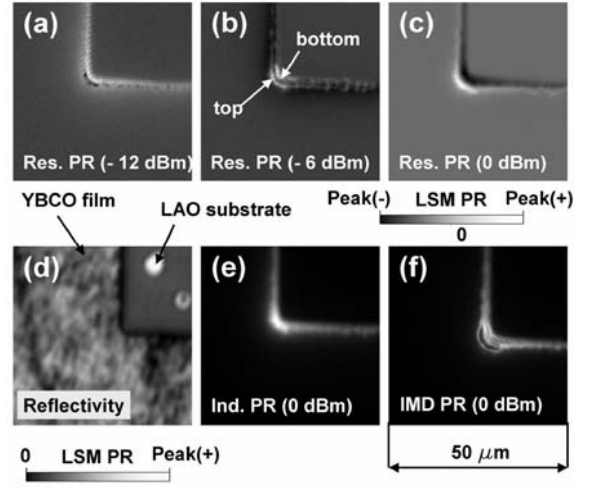

**Fig. 20**. The same 50x50 $\mu m^2$ area LSM scans obtained in different LSM imaging modes to show (a-c) power-dependent redistribution of the $\delta V_{RF}^{RES}(x,y)$ component, (d) optical map, (e) the $\delta V_{RF}^{IND}(x,y)$ distribution, and (f) nonlinear (IMD) responses.

(see Fig. 20,d). By comparison between the $\delta V_{RF}^{RES}(x,y)$ map, the $\delta V_{RF}^{IND}(x,y)$ distribution and the reflectivity LSM image, we established that the bottom of the YBCO strip becomes resistive at $P_{IN}$ = -24 dBm corresponding to $J_{PEAK}(x,y)$ = 2.7·$10^8$ A/$m^2$ at the inner edge of the right corner. The peak values of the $\delta V_{RF}^{RES}(x,y)$ photoresponse are the brightest areas in Fig. 20,a that form a broken line along the bottom edge. In this image obtained at $P_{IN}$ = -12 dBm, a small resistive state of the top edge starts to developed. Increasing $P_{IN}$ to – 6 dBm (Fig. 20,b) leads to destruction of superconductivity along the whole edge of YBCO strip. The most resistive areas in this situation coincide with both the top and bottom corners of the structure that are clearly visible in Fig. 20,b as two arcs created by $J_{PEAK}(x,y)$ = 1.1·$10^9$ A/$m^2$. At $P_{IN}$ = 0 dBm a wedge of the trapezoid is switched from the superconducting to the normal state due, in part, to a higher HF current density (4.5·$10^9$ A/$m^2$) at the bottom of the strip. The $\delta V_{RF}^{RES}(x,y)$ is disappearing in the areas with a normal conductivity as evident from Fig. 20,c. It should be emphasized that no visible geometry-related effect in the $\delta V_{RF}^{IND}(x,y)$ distribution (Fig.20,e) was detected for input powers between -50 dBm and +10 dBm at all temperatures from 78 K to $T_c$. Because of this, the inductive component of LSM PR may serve as a reliable indicator of the $J_{RF}(x,y)$ distribution and amplitude.

The power handling capability of typical HTS devices is limited by strong nonlinearity (NL) of the surface impedance in d-wave cuprates at high exciting HF field. This manifests itself as generation of spurious harmonics and intermodulation (IM) product distortion. The *global* third-order IM product was found to vary as the cube of $P_{IN}$ for $P_{RF}$ < -6 dBm, but began to saturate at higher power, partly motivating our study of the *local* IM PR, $\delta V_{IM}(x,y)$, the LSM imaged at the intermodulation tones $P_{2f1-f2}$ and $P_{2f2-f1}$. It is widely accepted that sources of this problem have both resistive and inductive origins, and arise locally as a result of high current densities $J_{RF}(x,y)$ nonuniformly distributed in the cross-section of the SC films. The lower limit of NLs can be attributed to the Nonlinear Meissner Effect (NLME), parameterized as a HF current (field) dependent magnetic penetration depth [66]:

$$\lambda_{eff}(T, J_{RF}) \cong \lambda_{eff}(T) * \left[ 1 + \frac{1}{2} \left( \frac{J_{RF}(x,y)}{J_{NL}(x,y)} \right)^2 \right] \quad (14)$$

Here $\lambda_{eff}(T)$ is the local value of the temperature dependent magnetic penetration depth and $J_{NL}$ is a material parameter that corresponds to the intrinsic depairing current density in HTS. Large current densities can be reached at the strip edge because of the non-uniform Meissner screening even at modest input power levels. For currents above $J_{NL}$, penetration of HF vortices, hysteresis, and thermal dissipation contribute to the NL of the HTS device. Grain boundaries, structural, geometrical and superconducting inhomogeneities generate additional resistive NL sources, reducing the power handling capability even further.

In this case, HF currents peaked at the right-angle inside corners of the transmission line strips play an essential role in determining the nonlinear behavior. Fig. 20,f shows the distribution of the $\delta V_{IM}(x,y)$ at $P_{IN}$ = 0 dBm, $f_{IM}$ = $2f_1$-$f_2$. The brightest areas in the picture correspond to the (positive) maximum of IM current density while the darkest one show the negative maximum of $\delta V_{IM}(x,y)$ signal. As can be seen, the positive peaks of $\delta V_{IM}(x,y)$ are strongly localized along the HTS strip edge not too far from the corner. The local value of $J_{RF}(x,y)$ in those positions reaches the value of critical current density of intergrain weak links. Appearance of the negative $\delta V_{IM}(x,y)$ between the positive peaks can be attributed to resistive (vortex-excited) NL component that generates the LSM PR of the opposite sign. Such behavior is possible when either Abrikosov ($\lambda$< thickness *d*) or Pearl ($\lambda$>*d*) vortices are generated by high current densities at the corner. These vibrating vortices have a tendency to move into the film under the influence of the Lorenz force with the frequency of the applied rf drive, and to create electrical field of opposite sign to the applied one. However, effects of heating in the HTS film by hot-spots of high rf current densities should not to be excluded from consideration as well. The lack of a theory for nonlinear LSM PR does not allow us to make a definitive conclusion.

By this expedient we have developed the LSM method for spatially-resolved investigation of correlation between (*i*) position of individual NL sources, (*ii*) distribution of rf current and (*iii*) location of resistive losses in HTS devices. It is shown that the dominant NL sources are generated mainly by peak current density at the HTS strip edges and is caused by nonuniform dissipation due to the normal component of $J_{RF}(x,y)$.

### 3.3. Two-beam arrangement

It was proven in previous works that the laser probe can locally manipulate the position of resistive domains [67] and light-sensitive parameters of a superconductor such as $J_c(x,y)$ [20], $\lambda_{eff}(x,y)$ [14], etc. [21]. This gives the means to solve the problem of re-arrangement of transport properties in HTS due to spatial inhomogeneities of the superconducting characteristics, of the material structure, and of the sample geometry and local defects of any sort. As an example, Fig. 21 shows LSM images visualizing the effect of re-distribution of magnetic flux flow channels in a YBCO thin film strip biased by dc current *I*=140 mA. Fig. 21,a illustrates the resistive LSM response of the structure when one artificial defect is present, and Fig. 21,b is the same but with two artificial defects. The spatial distribution of electric field amplitude *E*(*x*,*y*) in Fig. 21,a correlates with analytic and numerical calculations by Friesen and Gurevich [64] who have shown anomalous increase in density of the current paths near the HTS strip edge damaged with a linear defect. The $J_c(x,y)$ critical density region looks in the figure like a bright spot near the cut. It initiates entrance of the vortices which move along the channels imaged by bright strips on the background of a dark-looking superconducting state. The role of the isolating defect with a fixed position near the HTS strip edge is played by the laser cut denoted by the white arrow. Fig. 21,b presents the image of the resistive state obtained under the same conditions but spatially modified by an additional defect. The size, influence power and position of this defect were modeled by fitting the intensity of an additional (to the probing one) laser beam being focused into various points of the HTS strip. The change of the trajectory of magnetic flux flow is obviously seen in Fig. 21,b. It is caused by intermediate annihilation of vortex-antivortex pairs in the spot of the additional focused laser beam. A similar procedure of manipulating the local superconducting properties of an HTS structure can be apparently utilized to create more complicated zones of spatially-modulated order parameter in HTSs like, e.g. a vortex lattice. To do this, we suggest focusing the optical patterns onto the surface of the HTS structure by means of interference filters.

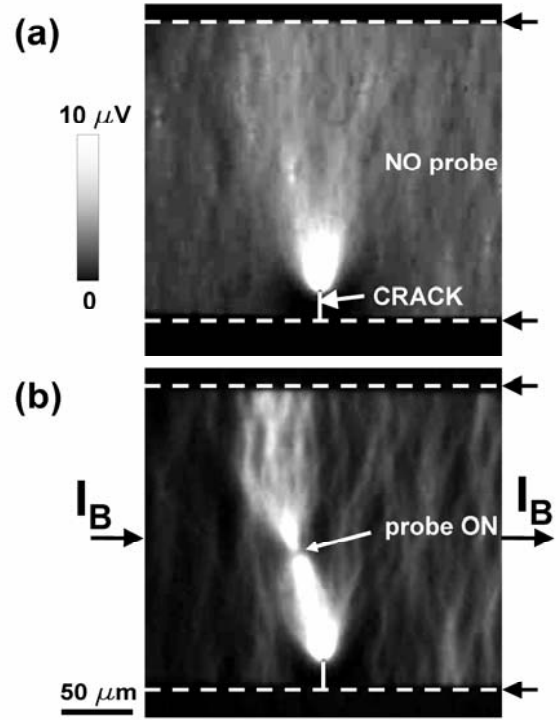

**Fig. 21**. 250x250 μm² 2-D LSM dc voltage image of the resistive state in a YBCO strip fragment induced by transport current *I*=140 mA at temperature *T*=88.7 K: (a) an artificial linear defect (crack) is present near the bottom edge, (b) the second laser beam generates additional inhomogeneity. The strip width is 200 μm, it is indicated by arrows and outlined by dashed lines.
 

## 4. CONCLUSION

This paper illustrates the potential of the LSM technique developed by the authors for non-contact, spatially-resolved probing of superconducting characteristics of HTS materials and devices. This method is capable of mapping optical, thermal, rf and dc electron transport properties and superconducting critical parameters of HTS samples directly in their operating state at $T<T_c$, with micron-scale spatial resolution. During scanning, different LSM contrast imaging modes are used to establish a spatial correlation between these diverse physical properties of the sample. Information of this sort can be used to solve technological, technical and physical problems associated with HTS materials. In addition, the newly developed option of the two-probe LSM device is demonstrated. It enables new and unique experiments on imaging local parameters of a superconductor under reversibly controllable manipulation of the topology of superconducting properties by additional spatially-patterned laser beams. This opens a new way of evolution of non-destructive testing of HTS devices and physical experiment techniques on studying spatially non-uniform electron transport and magnetic structures. Other interesting utilizations of LSM are nevertheless left aside because of limited room in the paper. However, some LSM potentialities implemented in other groups are worth noting. They are the combination of LSM with ellipsometric [68] or Raman [69] spectroscopy to spatially explore the HTS materials structure, and also the THz-band LSM for spatially-resolved studies of superconductor's gap features [70,71]. Among the promising but not yet demonstrated techniques involving the LSM are the two-frequency LSM with pumping and probing lasers to examine spatial variations of quasi-particle excitations [72], and relaxation of electron and phonon processes in HTSs distinguished by time scale hierarchy [73].

## ACKNOWLEDGEMENTS

We acknowledge valuable contributions from D. Abraimov (Applied Superconductivity Center, Madison, USA), A. Lukashenko (Erlangen University, Germany), and Stephen Remillard (Agile Devices, USA). This work has been supported in part by the NSF/GOALI DMR-0201261, the program «Nanosystems, nanomaterials, and nanotechnology» of the National Academy of Sciences of Ukraine, and a DFG Grant "Vortex matter in mesoscopic superconductors".